\documentclass[aps,prb,twocolumn,longbibliography,floatfix,10pt,superscriptaddress]{revtex4-2}
\usepackage[utf8]{inputenc}
\usepackage{mathrsfs}
\usepackage{natbib}
\usepackage{graphicx}
\usepackage{latexsym}
\usepackage{amssymb}
\usepackage{amsmath}
\usepackage{amsfonts}
\usepackage{gensymb}
\usepackage{bm}
\usepackage{hyperref}
\usepackage{braket}
\usepackage{physics}
\usepackage{bbold}
\usepackage[caption=false]{subfig}
\usepackage[dvipsnames]{xcolor}
\usepackage[normalem]{ulem}
\usepackage{siunitx}
\usepackage{mhchem}
\usepackage{multirow}
\usepackage{soul}

\usepackage{tikz-cd}
\usetikzlibrary{decorations.pathreplacing}

\usepackage{slashed}

\usepackage{float} 

\definecolor{myforestgreen}{RGB}{34,139,34}

\newcommand{\PRLsec}[1]{\emph{#1---}}

\newcommand{\supplementarysection}{%
 \setcounter{figure}{0}
 \let\oldthefigure\thefigure
 \renewcommand{\thefigure}{S\oldthefigure}
 \setcounter{section}{0}
 \let\oldthesection\thesection
 \renewcommand{\thesection}{S\oldthesection}
 \setcounter{equation}{0}
 \let\oldtheequation\theequation
 \renewcommand{\theequation}{S\oldtheequation}
 \setcounter{table}{0}
 \let\oldthetable\thetable
 \renewcommand{\thetable}{S\oldthetable}
}

\newenvironment{dfn}{{\vspace*{1ex} \noindent \bf Definition }}{\vspace*{1ex}}

	\newcommand{\beq}{\begin{eqnarray}}
	\newcommand{\eeq}{\end{eqnarray}}
	\newcommand{\bea}{\begin{eqnarray}\begin{aligned}}
	\newcommand{\eea}{\end{aligned}\end{eqnarray}}

	\renewcommand{\v}{{\bf v}}

\begin{document}

\title{Anyon molecules in fractional quantum Hall states}

\author{Taige Wang}
\affiliation{Department of Physics, University of California, Berkeley, CA 94720, USA \looseness=-2}
\affiliation{Department of Physics, Harvard University, Cambridge, MA 02138, USA \looseness=-2}
\affiliation{Materials Research Laboratory, Massachusetts Institute of Technology, Cambridge, MA 02139, USA \looseness=-2}

\author{Michael P. Zaletel}
\affiliation{Department of Physics, University of California, Berkeley, CA 94720, USA \looseness=-2}

\begin{abstract}
We use segment DMRG on the infinite cylinder to compute energies of charged excitations in gate-screened fractional quantum Hall states. For the $\nu=1/3$ Laughlin state, the $\nu=2/5$ Jain state, and the $\nu=5/2$ anti-Pfaffian state, we find screening can bind like-charged anyons into molecules, with a strong dependence on filling factor, gate distance, and fusion channel. In the Laughlin state, stable $\pm 2e/3$ molecules and larger clusters appear over a broad gate-distance window. The Jain state is molecular throughout the range we consider. In the anti-Pfaffian, binding is strongest on the hole side, where the charge-$e/2$ molecule prefers the $\psi$ fusion channel throughout the range we consider. In all three cases, screening suppresses long-range repulsion and exposes an intermediate-range attraction encoded in the oscillatory density tail of the fundamental anyon. We discuss consequences for addition spectra, interferometry, Wigner crystallization, anyon superconductivity, and entropy measurements.
\end{abstract}

\maketitle

\PRLsec{Introduction}
Fractional quantum Hall (FQH) states support anyons with fractional charge, fractional statistics, and topological ground-state structure fixed by long-wavelength field theory~\cite{Laughlin1983PRL,Arovas1984PRL}. That effective description does not fix the energetic hierarchy within a given charge sector. In a realistic screened device, do like-charged anyons remain isolated, or do they bind into molecules? This is a microscopic question.\footnote{An interesting precedent already exists at $\nu=1$, where sufficiently small Zeeman energy can favor a charge-two skyrmion over two charge-one skyrmions.~\cite{LilliehookEtAl1997}.}

The $\nu=1/3$ Laughlin state is the simplest setting. In the composite-fermion picture, the elementary hole and particle, with charges $e/3$ and $-e/3$, are the fundamental charged particles, while higher-charge excitations are built from weakly interacting collections of them~\cite{Jain1989CF,JainBook}. However, nearby metallic gates can render these residual interactions attractive at intermediate distances, potentially stabilizing anyonic molecules.

Here we compute thermodynamic fixed-charge excitation energies with segment DMRG on the infinite cylinder~\cite{ZaletelMongPollmann2013}. The method creates localized charged excitations in the center of a long cylinder and converges their energies with respect to circumference, segment length, and bond dimension. Using $L_y\gtrsim 18\,\ell_B$ and segment lengths above $60\,\ell_B$, we find no detectable drift in the relevant energy differences upon further enlargement. Throughout, the electron charge is $-e$, so holes carry positive charge and particles negative charge.

Our main result is that screening binds like-charged anyons, with strong state and fusion-channel dependence. In the $\nu=1/3$ Laughlin state, a broad window $d_g/\ell_B=O(1)$ stabilizes $\pm 2e/3$ molecules and larger clusters. The Jain $\nu=2/5$ state binds even more readily: in every sector we access, the lowest fixed-$Q$ state lies below the isolated-anyon threshold. At $\nu=5/2$, the iDMRG ground state is the anti-Pfaffian (APf), which also binds at small $d_g$, with pronounced particle-hole asymmetry and strong fusion-channel dependence in the charge-$e/2$ sector. The dilute defects remain the familiar $\pm e/3$, $\pm e/5$, and $\pm e/4$ anyons. Screening instead reorganizes the finite-density hierarchy.

The mechanism is common across all three states. An isolated anyon already carries an oscillatory density tail. Screening removes the long-range repulsion and exposes an intermediate-range attraction at a preferred separation, producing a finite-spacing molecule rather than a collapsed defect. Single-anyon probes should still resolve charges $e/3$, $e/5$, and $e/4$, while compressibility, capacitance, and addition spectra can be governed by clustered objects such as $2e/3$, $2e/5$, and $e/2$. In the APf state, the broad preference for the hole $\psi$ channel implies that finite-density entropy probes may not access the naive fusion degeneracy.

\begin{figure}[t]
\includegraphics[width=\columnwidth]{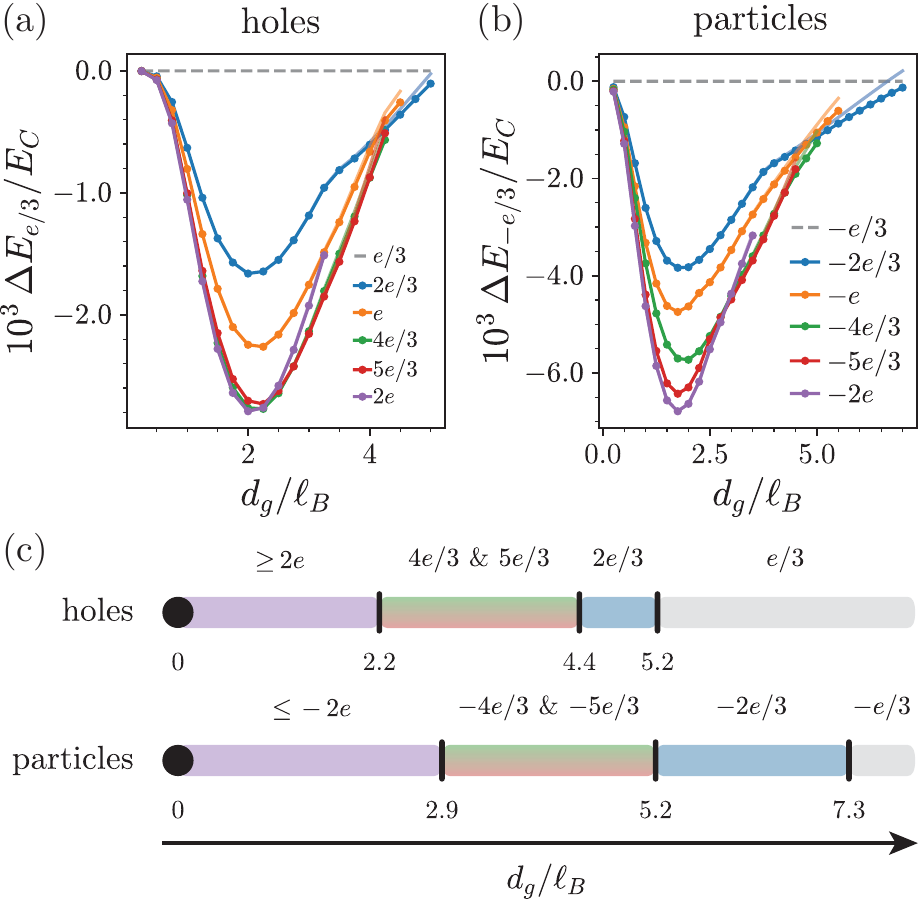}
\caption{
\textbf{Gate-tuned hierarchy of anyonic molecules in the $\nu=1/3$ Laughlin state.} (a),(b) Lowest fixed-$Q$ energies per elementary hole ($e/3$) and particle ($-e/3$), relative to isolated anyons, versus gate distance $d_g$. Faint lines show raw $L_y=18\ell_B$ data before the Madelung correction. We show only the centrally localized branch. Curves end when the variational minimum switches to a split-edge configuration. (c) Preferred added charge on the hole and particle sides. In the shaded regions, the splittings between $4e/3$ and $5e/3$, and between $-4e/3$ and $-5e/3$, are within numerical accuracy.}
\label{fig:pd}
\end{figure}

\PRLsec{Anyon binding}
We work in a single Landau level on the infinite cylinder. The projected Hamiltonian is
\begin{equation}
\hat H=\frac{1}{2A}\sum_{\mathbf q}V(\mathbf q)\rho_{-\mathbf q}\rho_{\mathbf q},
\label{eq:Hproj}
\end{equation}
where $\rho_{\mathbf q}$ is the Landau-level projected density operator. For symmetric top and bottom gates a distance $d_g$ from the two-dimensional electron gas, the screened interaction is
\begin{equation}
V(q)=2\pi E_C\ell_B\,\frac{\tanh(d_g q)}{q},
\label{eq:screenedV}
\end{equation}
with $q=|\mathbf q|$ and $E_C=e^2/(4\pi\epsilon \ell_B)$. It interpolates between bare Coulomb at large $d_g$ and a short-ranged interaction when $q d_g\ll 1$. In an anisotropic dielectric environment, such as graphene encapsulated by hBN, one rescales to $d_g^{\rm eff}=d_g\sqrt{\epsilon_{xy}/\epsilon_z}$. We focus on GaAs Landau levels without Landau-level mixing. Graphene requires only the corresponding interaction form factor.

We first obtain the uniform ground state using iDMRG on an infinite cylinder, yielding topologically degenerate ground states labeled by anyonic flux $a$. Charged excitations are then computed with segment DMRG on the same cylinder. A finite central segment is inserted between two semi-infinite ground-state MPSs carrying topological flux $[l,r]$ to the left and right. The boundary conditions $[l,r]$ force a prescribed anyon $a$ into the segment according to the fusion rule $r = l \times a$. The proper choice of boundary condition for each state is summarized in the Appendix. With $Q$ and $[l,r]$ fixed, the optimization yields the lowest energy $E_Q$ in each multi-anyon sector, measured relative to the vacuum.

To account for the leading finite-$L_y$ correction, we subtract the gate-screened Madelung interaction between an anyon and its periodic images. The Madelung formula and convergence analysis in $L_y$, MPS bond dimension $\chi$, and segment length $L_x$ are given in the Appendix. Unless noted otherwise, the main-text data use $L_y=18\,\ell_B$, $\chi=4800$, $L_x\approx 67\,\ell_B$, and $d_g\le 10\,\ell_B$. Because charged anyons obey a continuous magnetic algebra, their dispersion is flat, so localizing an anyon into the segment carries no extra kinetic cost. When a nominal charge-$Q$ molecule becomes unstable, the variational optimum jumps to a configuration with constituent charges at opposite segment ends, with residual interaction of order $V(L_x)(e^*/e)^2$. In that regime, energies drift with $L_x$, so we plot only the range of $d_g$ where no such drift is observed.

To quantify binding, we compare each fixed-charge sector with well-separated elementary anyons, which keeps the analysis independent of chemical potential. For a state with elementary charge magnitude $e^*=e/3$, $e/5$, or $e/4$, we write $Q = \pm n e^*$ and define
\begin{equation}
\Delta E_Q\equiv \frac{E_Q}{n}-E_{\mathrm{sgn}(Q)e^*},
\label{eq:binding}
\end{equation}
so that $\Delta E_Q<0$ means the charge-$Q$ sector lies below $n$ isolated elementary anyons of the same sign. Since $\Delta E_Q$ is defined per elementary anyon, the total binding gain is $E_{\mathrm{bind}} \equiv -n\Delta E_Q$. More generally, if $Q= m e^*$ and $Q'= n e^*$ have the same sign, then $\Delta E_Q<\Delta E_{Q'}$ implies $E_Q/|m|<E_{Q'}/|n|$: at fixed added charge it is cheaper to assemble the system from charge-$Q$ molecules than from charge-$Q'$ molecules. This criterion decides, for example, whether charge $4e^*$ prefers a tetramer or two dimers.

\begin{figure}[t]
\includegraphics[width=\columnwidth]{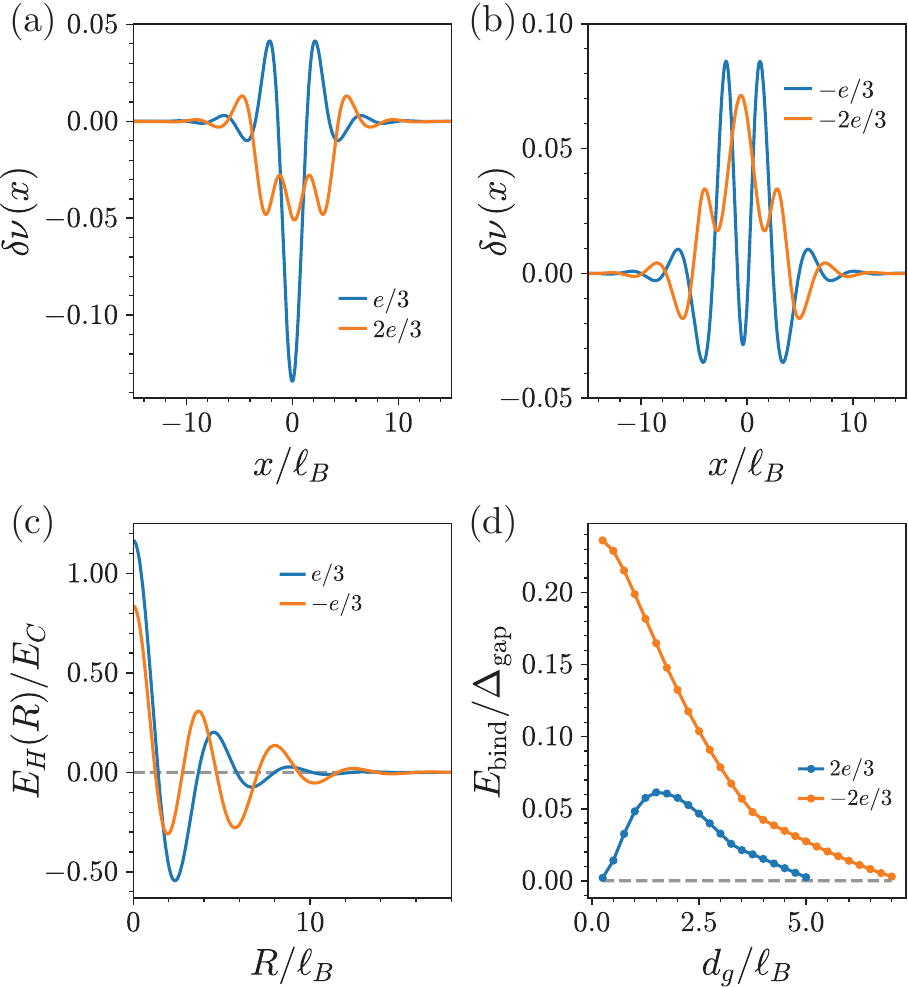}
\caption{\textbf{Oscillatory density tails drive binding.} (a),(b) Line cuts of the density modulation $\delta\nu(x)$ for particle sectors $Q=-e/3,-2e/3$ and hole sectors $Q=e/3,2e/3$ at $\nu=1/3$, $L_y=16\,\ell_B$, and $d_g=\ell_B$. (c) Hartree energy illustration obtained by displacing the single-anyon density textures in (a),(b) along $x$. (d) Binding energies $E_{\mathrm{bind}} \equiv - 2 \Delta E_{\pm 2e/3}$ normalized by the charge gap $\Delta_{\mathrm{gap}}$. The $2e/3$ hole molecule loses binding as $d_g\to0$, while the $-2e/3$ particle molecule retains a substantial fraction of the gap scale.}
\label{density}
\end{figure}

Fig.~\ref{fig:pd} shows the binding energies in the $\nu=1/3$ Laughlin state. On the hole side, the preferred added charge evolves from isolated $e/3$ holes to $2e/3$ molecules and then to larger clusters as screening is strengthened. The particle side shows the same sequence, with binding stable to larger gate distances. In the intermediate regime, the $4e/3$ and $5e/3$ hole sectors, and likewise the $-4e/3$ and $-5e/3$ particle sectors, are nearly degenerate within numerical accuracy [Fig.~\ref{fig:pd}(c)]. Over a broad window $d_g/\ell_B=O(1)$, the lowest-energy way to add finite charge is therefore an anyon molecule rather than a gas of isolated anyons. For $Q=\pm2e/3$, the total binding gain reaches $\sim10^{-2}E_C$, or about $10\%$ to $20\%$ of the charge gap [Fig.~\ref{density}(d)]. The drift toward larger preferred $|Q|$ as the interaction range is reduced suggests an incipient clustering instability, a finite-charge precursor of the tendency of the doped $\nu=1/3$ liquid to nucleate nearby incompressible descendants.

The short-range limit is clearest in Fig.~\ref{density}(d). Because Fig.~\ref{fig:pd} is plotted in bare Coulomb units $E_C$, reducing $d_g$ trivially softens the overall scale. Normalizing the $Q=\pm2e/3$ binding by the charge gap removes this rescaling. The $2e/3$ hole molecule then loses binding as $d_g\to0$, whereas the $-2e/3$ particle molecule retains a substantial fraction of the gap scale. This is the hard-core limit of the Laughlin problem. Holes are exact zero modes of the contact interaction and become asymptotically noninteracting~\cite{TrugmanKivelson1985}, while particles have no analogous protection. The details differ in the Jain and APf states, but the operative mechanism is the same intermediate-range attraction revealed by screening.

\begin{figure*}[t]
\includegraphics[width=2\columnwidth]{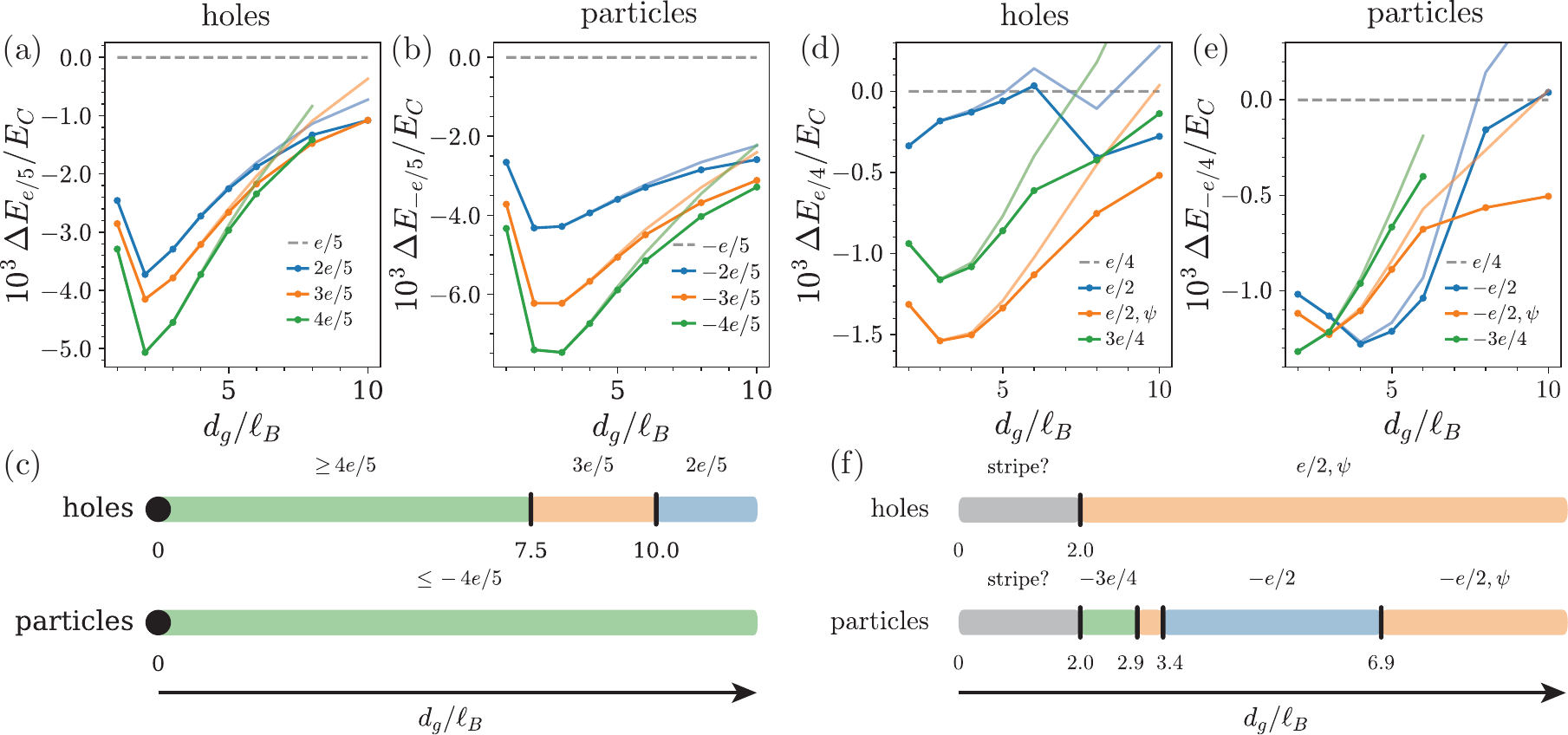}
\caption{\textbf{Gate-tuned hierarchy in the Jain and APf states.} (a),(b) Lowest fixed-$Q$ energies per elementary hole ($e/5$) and particle ($-e/5$) in the Jain $\nu=2/5$ state, relative to isolated anyons, versus gate distance $d_g$. (c) Preferred added charge on the hole and particle sides. (d),(e) Same for the APf state at $\nu=5/2$, now relative to isolated holes ($e/4$) and particles ($-e/4$). The $\pm e/2$ sectors are shown in both the $1$ and $\psi$ channels. (f) Preferred added charge in the two APf sectors. When a charge-$\pm e/2$ pair is favored, its fusion channel is indicated. Faint lines show raw $L_y=18\,\ell_B$ data before the Madelung correction. Only the central-localized branch is shown.}
\label{fig:jain_pf}
\end{figure*}

\PRLsec{Binding at $\nu = 2/5$ and $5/2$.}
We repeated the same fixed-charge analysis for the Jain $\nu=2/5$ state and the APf state at $\nu=5/2$, defining the binding energy relative to elementary anyons of charge $e/5$ and $e/4$. Fig.~\ref{fig:jain_pf} shows that screening again lowers multi-anyon sectors below the isolated-anyon threshold over a substantial gate-distance window. Like-charged binding is therefore not peculiar to the Laughlin state. Across these representative Abelian and non-Abelian phases, screening favors finite-density clustering, although the window, binding scale, and preferred topological sector remain strongly state dependent.

For the Abelian $\nu=2/5$ state, the evidence is especially clear. On both particle and hole sides, the $\pm2e/5$, $\pm3e/5$, and $\pm4e/5$ sectors lie below the isolated-anyon threshold throughout the full screening range shown. The Jain state is thus molecular across the experimentally relevant window we access, with total binding gains of a few $\times10^{-3}E_C$ on the particle side and of order $10^{-3}E_C$ on the hole side.

The APf state is richer because the charge-$e/2$ sector has two fusion channels, which we denote $e/2$ and $e/2,\psi$. On the particle side, the preferred channel changes with screening: the $e/2,\psi$ sector is lowest at short gate distance, but the $e/2$ sector overtakes it at larger $d_g$, while at still smaller $d_g$ a competing stripe instability may intervene~\cite{RezayiHaldane2000}. On the hole side, binding is stronger and more robust, and the $e/2,\psi$ sector remains well below $e/2$ throughout the bound-$e/2$ regime. The APf state therefore not only binds like-charged anyons, but also selects a definite fusion channel at finite density. Numerically, however, it is also more delicate than the Laughlin and Jain cases: even within a fixed charge sector, several molecular branches can compete, producing the kinks visible in the APf curves. As we discuss in the Appendix, some of these candidate molecules have diameters as large as $\sim 25\,\ell_B$, we therefore use segment lengths up to $L_x\simeq 112\,\ell_B$. Although the apparent hierarchy extends to large $d_g$, the corresponding binding energies are already below $10^{-3}E_C$, much smaller than in the Laughlin and Jain states. For the Pf state, the same results apply with particles and holes exchanged.

\PRLsec{Microscopic origin of binding}
The same microscopic mechanism underlies binding in all three hierarchies. As shown in Figs.~\ref{density}(a),(b), both quasiparticles and quasiholes in the Laughlin state exhibit oscillatory density profiles, with alternating positive and negative modulations away from the core. Once the interaction is screened to be sufficiently short-ranged, these oscillations make it favorable for two like-charged anyons to sit a few $\ell_B$ apart, so that a peak in one profile aligns with a valley in the other. The $Q=\pm 2e/3$ excitation can thus be viewed as a bound dimer of two $Q=\pm e/3$ anyons at a preferred separation.

To make the mechanism intuitive, Fig.~\ref{density}(c) shows a Hartree-like overlap energy built from single-anyon density profiles after displacing one profile relative to the other along $x$. Denoting the charge-density modulation of a single anyon by $\delta\rho(\mathbf r)$, we approximate two identical anyons separated by $R\hat x$ as rigid textures and evaluate the Hartree term,
\begin{equation}
E_H(R)=
\int d\mathbf r\,d \mathbf r'\,\delta\rho_a(\mathbf r)\,
V(\mathbf r-\mathbf r')\,
\delta\rho_b(\mathbf r'-R\hat x).
\label{eq:hartree}
\end{equation}
We do not expect $E_H$ to agree quantitatively with the binding energy, since non-Hartree effects dominate outside the thin-torus limit, but it captures the mechanism. For a non-oscillatory charge lump, the overlap would remain repulsive. The oscillatory anyon profile changes that conclusion. Once screening suppresses the long-range tail, the first few density lobes dominate, and the energy is lowered when a peak of one anyon aligns with a valley of the other. $E_H$ turns negative over a finite intermediate window of $R$. Screening does not create a new interaction channel, but rather it reveals an attraction already encoded in the oscillatory single-anyon density.

\PRLsec{Experimental implications}
Our results predict binding in existing gated 2DEGs. Since $\ell_B=25.7/\sqrt{B[\mathrm{T}]}\,\mathrm{nm}$, the regime $d_g^{\rm eff}\sim 5$ to $10\,\ell_B$ corresponds to $40$ to $\SI{80}{nm}$ at $B=10\,\mathrm{T}$~\cite{LaturiaEtAl2018,NakamuraEtAl2019AB,CohenEtAl2023,WerkmeisterEtAl2024}. For hBN-encapsulated graphene, dielectric anisotropy gives $d_g^{\rm eff}\approx1.5d_g$, and binding should be prevalent in typical $d_g \sim 30$--$\SI{40}{nm}$ heterostructures. 
In Chern bands in moir\'e systems, a phenomenological translation of our results is obtained by replacing $\ell_B \to \ell_M=\sqrt{A_M/2\pi}\approx0.37a_M$, which yields comparable scales~\cite{ZengEtAl2023MoTe2,AronsonEtAl2025,WatersEtAl2025Pentalayer,LuEtAl2024FQAH}. Binding energies of $10^{-3}$ to $10^{-2}E_C$ correspond to sub-kelvin temperatures in GaAs and kelvin-scale temperatures in graphene and moir\'e platforms.

Elucidating the experimental consequences of binding requires an understanding of thermodynamics in addition to energetics. 
As the chemical potential $\mu$ is tuned away from the center of the FQH gap, charge-$Q$ anyons begin to occupy an effective Landau level with flux density $(|Q|/e)/(2\pi \ell_B^2)$~\footnote{Fractional exclusion statistics and residual interactions between anyons are not relevant in the dilute limit we consider.}.
In the dilute, non-degenerate limit the density of monomers and dimers is thus
\begin{align}
n_{e/3}&= \frac{1}{6\pi \ell_B^2} e^{-\frac{E_{e/3}-\mu/3}{k_B T}} = 
\frac{1}{6\pi \ell_B^2} e^{-\frac{\Delta_{\rm gap} / 2 - \delta \mu/3}{k_B T}} \\
n_{2e/3}&= \frac{1}{3\pi \ell_B^2} e^{-\frac{E_{2e/3}-2\mu/3}{k_B T}} = \frac{1}{3\pi \ell_B^2} e^{-\frac{\Delta_{\rm gap} - E_{\rm bind} -2\delta \mu/3}{k_B T}}.
\label{eq:anyon_densities}
\end{align}
where $\delta \mu$ is measured relative to the center of the gap. 
Their relative abundance is
\begin{equation}
\frac{n_{e/3}}{n_{2e/3}}
=
\frac12
e^{\frac{\Delta_{\rm gap}/2 - E_{\mathrm{bind}} -\delta\mu/3 }{k_B T}},
\label{eq:saha_main}
\end{equation}
Deep in the gap, $\Delta_{\rm gap}/2 - E_{\mathrm{bind}} >\delta\mu/3$: both densities vanish as $T\to0$, but monomers predominate. This is the regime relevant for activated transport~\cite{PolyakovShklovskii1994}, so the activation gap remains the monomer gap $\Delta_{\rm gap}=E_{e/3}+E_{-e/3}$ even when molecules are preferred energetically.
The physics here is analogous to Saha ionization~\cite{Saha1920Ionization}: at low density, molecules dissociate because monomers have more positional entropy. 

In contrast, measurements of the thermodynamic gap, defined as the width $\Delta \mu$ of the incompressible region in the $T \to 0$ limit, are sensitive to binding. 
 As $\delta \mu/3$ approaches the gap edge, according to Eq.~\eqref{eq:anyon_densities} dimers appear at $\delta \mu / 3 = \Delta_{\rm gap}/2 - E_{\rm bind}/2$, before the monomer edge at $\delta \mu / 3 = \Delta_{\rm gap}/2$. Thus, for $k_B T\lesssim E_{\rm bind} / 2$, the width $\Delta \mu$ is set by the energy of dimers, or more generally the particle-like and hole-like $Q_\pm^\ast$ with the lowest energy per unit charge,
\begin{equation}
\Delta \mu = \sum_{\pm}\frac{e}{|Q_\pm^\ast|}E_{Q_\pm^\ast},
\end{equation}
which is reduced relative to the unbound value $3\Delta_{\rm gap}$.

As $\delta \mu$ is increased beyond the gap, one enters the finite-density regime probed in measurements of negative compressibility~\cite{EisensteinEtAl1992NegativeCompressibility}. Defining $T_{\rm bind}$ by the condition $n_{\pm e/3}=2n_{\pm 2e/3}$, we find 
\begin{equation}
\label{eq:Tbind}
k_B T_{\rm bind}
=
-\frac{E_{\rm bind}}{\ln(18|\delta\nu|)}.
\end{equation}
where $\delta \nu$ is the deviation in the electron filling factor. 
For the experimentally relevant window $|\delta\nu|\sim 0.01$ to $0.02$, this gives $k_B T_{\rm bind}\approx0.8\,E_{\rm bind}$. Below $T_{\rm bind}$, the predominant carriers are molecules.
At sufficiently low temperature, these molecules will form a pinned Wigner crystal whose lattice constant reveals the molecular charge $Q$~\cite{AssoulineEtAl2026Entropy}:
\begin{equation}
n_Q=\frac{e|\delta\nu|}{2\pi \ell_B^2 |Q|},
\qquad
a_Q=\sqrt{\frac{2}{\sqrt{3}\,n_Q}}
=\ell_B\sqrt{\frac{4\pi}{\sqrt{3}}\frac{|Q|/e}{|\delta\nu|}}
\end{equation}
(assuming a triangular lattice). At fixed $|\delta\nu|$, a $Q=2e/3$ crystal has a lattice spacing larger by a factor of $\sqrt{2}$ than a conventional $e/3$ crystal. STM could image this spacing, while local charge sensing could determine the cluster charge~\cite{HuEtAl2025,ChiuEtAl2025,AssoulineEtAl2024}. Because the interaction energy per cluster scales as $Q^2/a_Q\propto Q^{3/2}\sqrt{|\delta\nu|}$, the filling dependence of the chemical potential is modified accordingly~\cite{EisensteinEtAl1992NegativeCompressibility,Cooper2009,YangHalperin2009Thermopower}.

The possibility of bound $e/2$ molecules in the Pf/APf state is especially relevant to proposals for thermodynamic detection of non-Abelian entropy~\cite{Cooper2009,AssoulineEtAl2026Entropy}, which rely on charges entering as well-separated non-Abelian $e/4$ anyons at finite density. On the hole side, Fig.~\ref{fig:jain_pf} shows that the $e/2,\psi$ dimer lies well below both the $e/4$ monomer and the $e/2$ dimer over a broad window of gate distances. Finite-density doping therefore produces Abelian bound pairs in the preferred $\psi$ channel rather than a macroscopically degenerate gas of $e/4$ holes, quenching the predicted non-Abelian entropy. Thermally repopulating the $1$ channel would require temperatures of order $10^{-3}E_C$, but in this regime non-topological contributions to the entropy from the anyons' positional degree of freedom swamp the topological contribution. Finite-density non-Abelian entropy is therefore difficult to access on the hole side. The particle side is more favorable, since the $e/2$ and $e/2,\psi$ sectors are close in energy, effectively restoring the non-Abelian degeneracy, while binding is lost at experimentally reasonable gate distances.

The charging of a quantum dot, antidot, or impurity instead probes a discrete-charging limit. We focus on dots which are significantly larger than individual quasiparticles and which contain a few separated charges, so as to behave like a compressible puddle. 
Then over a small occupancy window where the dot's total capacitance $C$ and lever arm $\alpha$ can be treated as constant, the energy of $n$ trapped $-e/3$ anyons at dot-gate voltage $V_g$ is approximated as 
\begin{equation}
E_n(V_g)\simeq \frac{((n - n_0) e/3)^2}{2C}-\frac{e}{3}\alpha V_g\,n
-\Big\lfloor\frac n2\Big\rfloor E_{\rm bind},
\end{equation}
where $E_{\rm bind}>0$ lowers the energy of \emph{each} bound pair and $n_0$ is a reference occupation. The loading threshold $V_g^{(n\to n+1)}$ is the gate voltage at which the ground state switches from $n$ to $n+1$ trapped anyons. Without binding, charging events are nearly evenly spaced at $\frac{e}{3} \alpha \Delta V_g = e^2 / 9 C$. With binding, the spacings alternate,
\begin{equation}
\frac{e}{3}\alpha\,\Delta V_g^{\rm even,odd}
=
\frac{e^2}{9C} \pm E_{\rm bind},
\end{equation}
with the upper (lower) sign for even (odd) occupancies. 
This alternating structure should be visible in Coulomb-diamond measurements of gate-defined dots or antidots near a quantum point contact:
\begin{equation}
\begin{centering}
\begin{tikzpicture}[x=0.7cm,y=0.6cm,thick]
 \def\we{3.2}
 \def\wo{1.6}
 \def\he{1.5}
 \def\ho{0.8}

 \draw
  (0,0) --
  ({\we/2},\he) --
  (\we,0) --
  ({\we+\wo/2},\ho) --
  ({\we+\wo},0) --
  ({\we+\wo+\we/2},\he) --
  ({2*\we+\wo},0);

 \draw
  (0,0) --
  ({\we/2},-\he) --
  (\we,0) --
  ({\we+\wo/2},-\ho) --
  ({\we+\wo},0) --
  ({\we+\wo+\we/2},-\he) --
  ({2*\we+\wo},0);

 \node at ({\we/2},0) {$n$};
 \node at ({\we+\wo/2},0) {$n\!+\!1$};
 \node at ({\we+\wo+\we/2},0) {$n\!+\!2$};

 \draw[->] (-0.9,-2.0) -- (-0.9,2.0) node[above] {$V_{\rm bias}$};

 \draw[|<->|] (0,-1.95) -- node[below=5pt] {$\Delta V_g^{\rm even}$} (\we,-1.95);

 \draw[<->] (\we,-1.95) -- node[below=5pt] {$\Delta V_g^{\rm odd}$} ({\we+\wo},-1.95);

 \draw[|<->|] (\we+\wo,-1.95) -- node[below=5pt] {$\Delta V_g^{\rm even}$} ({\we+\wo+\we},-1.95);

 \draw[<->] ({\we+\wo/2},\ho) --
        node[right=2pt] {$E_{\mathrm{bind}}$}
        ({\we+\wo/2},\he);

 \draw[dashed] ({\wo},\he) -- ({\wo+\we+\wo},\he);
\end{tikzpicture}
\end{centering}
\end{equation}
Once $E_{\rm bind}>e^2/9C$, odd occupancies are skipped and the ground-state staircase becomes even-only. 
Choosing typical parameters for a graphene heterostructure at $B = \SI{14}{T}$, the crossover $E_{\rm bind}=e^2/9C$ occurs at a dot area of order $A\approx (\SI{230}{nm})^2$~\footnote{Estimated from $C \simeq 2\epsilon_z A/d_g$ using $d_g=\SI{40}{nm}$, $\epsilon_z=3$, $\epsilon_{xy}=6.6$, $B=\SI{14}{T}$, and $E_{\rm bind}=0.004E_C$.}.
Recent experiments have already observed analogous even-odd effect in much larger antidot devices, although those structures lie in the mesoscopic regime~\cite{DiLucaEtAl2025FractionalCoulombmeter}. For still larger structures, such as the bulk of an interferometer, charge will enter predominantly as dimers at sufficiently low temperature [Eq.~\eqref{eq:Tbind}].

Interferometry can thus be affected if molecules, rather than elementary anyons, are added to the bulk as a gate voltage or magnetic field is swept. Whether this occurs again depends on whether $E_{\rm bind}$ exceeds the charging energy of the relevant compressible puddle or impurity, and therefore on device details. 
In the standard theory of a $1/3$ Fabry-P\'erot interferometer~\cite{HalperinEtAl2011,RosenowHalperin2007}, the phase jump produced by the entrance of a charge-$Q$ anyon into the bulk of the interferometer is
\begin{equation}
\begin{gathered}
\Delta\theta_{\rm obs}(Q)\simeq \Delta\theta_{\rm stat}(Q)+\delta\theta_{\rm BE}(Q),\\
\Delta\theta_{\rm stat}(Q)=2\pi\frac{Q}{e}\ \ ({\rm mod}\ 2\pi),
\qquad
\delta\theta_{\rm BE}(Q)\propto -\frac{K_{IL}}{K_I}\frac{Q}{e},
\end{gathered}
\label{eq:phasejump}
\end{equation}
where $\Delta\theta_{\rm stat}$ is the statistical phase and $\delta\theta_{\rm BE}$ is a nonuniversal bulk-edge coupling correction determined by the electrostatic parameter $K_{IL}/K_I > 0$~\cite{HalperinEtAl2011}. Because the bulk-edge term always shifts the phase jump toward zero~\cite{NakamuraEtAl2022}, a single $e/3$ event cannot produce a slip larger than $2\pi/3$. In contrast, a $Q=2e/3$ event carries a bare statistical phase of $4\pi/3$ modulo $2\pi$. Without bulk-edge coupling, this is indistinguishable from the entry of a $Q=-e/3$ quasiparticle, and could therefore be mistaken for the ``wrong'' polarity entering the interferometer~\cite{SamuelsonEtAl2026Slow,HenzingerEtAl2026Control}. Weak bulk-edge coupling resolves this ambiguity: a $2e/3$ event produces a slip between $2\pi/3$ and $4\pi/3$, while a $-e/3$ event produces one between $4\pi/3$ and $2\pi$. There is evidence for the former in a recent graphene interferometry experiment~\cite{SamuelsonEtAl2026Slow}.

Although our calculation concerns bulk anyons rather than edge dynamics, it may still help motivate renewed efforts to understand earlier edge-tunnelling and interferometric experiments, in which the effective tunnelling charge on hole-conjugate edges often appeared to be bunched, for example at $2e/3$, rather than equal to the minimal quasiparticle charge~\cite{BidEtAl2009,GhoshEtAl2025}. Recent photo-assisted shot-noise and Hong-Ou-Mandel measurements at $\nu=2/3$, however, recover an $e/3$ tunnelling charge and underscore the role of neutral modes~\cite{DeEtAl2025}, so a dedicated edge calculation is needed.

More broadly, the gate-tuned molecular regime realizes the hierarchy invoked in earlier theories of anyon superconductivity~\cite{Laughlin1988PRL,Laughlin1988Science,HalperinAnyonSC1992,FisherLeeAnyonSC1991}. When the lowest-energy doped excitation is a $2e/3$ molecule rather than an isolated $e/3$ anyon, the natural route out of the Laughlin state is to dope these objects first. This helps explain why superconductivity can readily emerge: single-electron excitations remain gapped, while charge-$2e$ Cooper-pair correlations can become long-ranged. Topological quantum field theory indeed shows that when charge enters most cheaply through bound anyon pairs, doping or kinetically melting the Laughlin liquid favors an Abelian chiral superconductor with $c_-=3$~\cite{ShiSenthil2025,PichlerEtAl2025,WangZaletel2025}. The phases favored when the preferred cluster charge exceeds $2e/3$ remain an open question.

\paragraph*{Note added.}
As this work was being completed, we became aware of Refs.~\cite{XuEtAl2025AnyonClusters,GattuJain2025MolecularAnyons,LiNosovWangKhalaf2026BoundStates}. Together with our data, they strengthen the evidence that screening favors like-charged clustering in FQH fluids. Ref.~\onlinecite{XuEtAl2025AnyonClusters} found Laughlin binding in exact diagonalization on the sphere and studied clustering tendencies in the Moore-Read zero-mode manifold. Ref.~\onlinecite{LiNosovWangKhalaf2026BoundStates} used geometric quantization at much larger sizes and showed that finite-size effects substantially shift the critical screening length. Ref.~\onlinecite{GattuJain2025MolecularAnyons} suggested clustering at $\nu=2/5$ and $\nu=5/2$ from trial states. Our results complement these works by incorporating experimentally relevant gate screening and by determining thermodynamic binding energies with minimal finite-size effects and no further approximations.

\textbf{Acknowledgments.} We thank Eslam Khalaf, T. Senthil, and Ashvin Vishwanath for helpful discussions. We thank Qingchen Li, Pavel A. Nosov, and Eslam Khalaf for collaboration on a related project. 
M.Z. was funded by the U.S. Department of Energy, Office of Science, Office of Basic Energy Sciences, Materials Sciences and Engineering Division under Contract No. DE-AC02-05-CH11231 (Theory of Materials program KC2301). T.W. is grateful for the support by the Harvard Quantum Initiative Fellowship and the Simons Collaboration on Ultra-Quantum Matter, which is a grant from the Simons Foundation (Grant No. 651440). This research uses the Lawrencium computational cluster provided by the Lawrence Berkeley National Laboratory (Supported by the U.S. Department of Energy, Office of Basic Energy Sciences under Contract No. DE-AC02-05-CH11231).

\bibliography{main}

\begin{thebibliography}{47}%
\makeatletter
\providecommand \@ifxundefined [1]{%
 \@ifx{#1\undefined}
}%
\providecommand \@ifnum [1]{%
 \ifnum #1\expandafter \@firstoftwo
 \else \expandafter \@secondoftwo
 \fi
}%
\providecommand \@ifx [1]{%
 \ifx #1\expandafter \@firstoftwo
 \else \expandafter \@secondoftwo
 \fi
}%
\providecommand \natexlab [1]{#1}%
\providecommand \enquote  [1]{``#1''}%
\providecommand \bibnamefont  [1]{#1}%
\providecommand \bibfnamefont [1]{#1}%
\providecommand \citenamefont [1]{#1}%
\providecommand \href@noop [0]{\@secondoftwo}%
\providecommand \href [0]{\begingroup \@sanitize@url \@href}%
\providecommand \@href[1]{\@@startlink{#1}\@@href}%
\providecommand \@@href[1]{\endgroup#1\@@endlink}%
\providecommand \@sanitize@url [0]{\catcode `\\12\catcode `\$12\catcode `\&12\catcode `\#12\catcode `\^12\catcode `\_12\catcode `\%12\relax}%
\providecommand \@@startlink[1]{}%
\providecommand \@@endlink[0]{}%
\providecommand \url  [0]{\begingroup\@sanitize@url \@url }%
\providecommand \@url [1]{\endgroup\@href {#1}{\urlprefix }}%
\providecommand \urlprefix  [0]{URL }%
\providecommand \Eprint [0]{\href }%
\providecommand \doibase [0]{https://doi.org/}%
\providecommand \selectlanguage [0]{\@gobble}%
\providecommand \bibinfo  [0]{\@secondoftwo}%
\providecommand \bibfield  [0]{\@secondoftwo}%
\providecommand \translation [1]{[#1]}%
\providecommand \BibitemOpen [0]{}%
\providecommand \bibitemStop [0]{}%
\providecommand \bibitemNoStop [0]{.\EOS\space}%
\providecommand \EOS [0]{\spacefactor3000\relax}%
\providecommand \BibitemShut  [1]{\csname bibitem#1\endcsname}%
\let\auto@bib@innerbib\@empty
\bibitem [{\citenamefont {Laughlin}(1983)}]{Laughlin1983PRL}%
  \BibitemOpen
  \bibfield  {author} {\bibinfo {author} {\bibfnamefont {R.~B.}\ \bibnamefont {Laughlin}},\ }\bibfield  {title} {\bibinfo {title} {Anomalous quantum hall effect: An incompressible quantum fluid with fractionally charged excitations},\ }\href {https://doi.org/10.1103/PhysRevLett.50.1395} {\bibfield  {journal} {\bibinfo  {journal} {Phys. Rev. Lett.}\ }\textbf {\bibinfo {volume} {50}},\ \bibinfo {pages} {1395} (\bibinfo {year} {1983})}\BibitemShut {NoStop}%
\bibitem [{\citenamefont {Arovas}\ \emph {et~al.}(1984)\citenamefont {Arovas}, \citenamefont {Schrieffer},\ and\ \citenamefont {Wilczek}}]{Arovas1984PRL}%
  \BibitemOpen
  \bibfield  {author} {\bibinfo {author} {\bibfnamefont {D.}~\bibnamefont {Arovas}}, \bibinfo {author} {\bibfnamefont {J.~R.}\ \bibnamefont {Schrieffer}},\ and\ \bibinfo {author} {\bibfnamefont {F.}~\bibnamefont {Wilczek}},\ }\bibfield  {title} {\bibinfo {title} {Fractional statistics and the quantum hall effect},\ }\href {https://doi.org/10.1103/PhysRevLett.53.722} {\bibfield  {journal} {\bibinfo  {journal} {Phys. Rev. Lett.}\ }\textbf {\bibinfo {volume} {53}},\ \bibinfo {pages} {722} (\bibinfo {year} {1984})}\BibitemShut {NoStop}%
\bibitem [{Note1()}]{Note1}%
  \BibitemOpen
  \bibinfo {note} {An interesting precedent already exists at $\nu =1$, where sufficiently small Zeeman energy can favor a charge-two skyrmion over two charge-one skyrmions.~\cite {LilliehookEtAl1997}.}\BibitemShut {Stop}%
\bibitem [{\citenamefont {Jain}(1989)}]{Jain1989CF}%
  \BibitemOpen
  \bibfield  {author} {\bibinfo {author} {\bibfnamefont {J.~K.}\ \bibnamefont {Jain}},\ }\bibfield  {title} {\bibinfo {title} {Composite-fermion approach for the fractional quantum hall effect},\ }\href {https://doi.org/10.1103/PhysRevLett.63.199} {\bibfield  {journal} {\bibinfo  {journal} {Phys. Rev. Lett.}\ }\textbf {\bibinfo {volume} {63}},\ \bibinfo {pages} {199} (\bibinfo {year} {1989})}\BibitemShut {NoStop}%
\bibitem [{\citenamefont {Jain}(2007)}]{JainBook}%
  \BibitemOpen
  \bibfield  {author} {\bibinfo {author} {\bibfnamefont {J.~K.}\ \bibnamefont {Jain}},\ }\href {https://doi.org/10.1017/CBO9780511607561} {\emph {\bibinfo {title} {Composite Fermions}}}\ (\bibinfo  {publisher} {Cambridge University Press},\ \bibinfo {address} {Cambridge},\ \bibinfo {year} {2007})\BibitemShut {NoStop}%
\bibitem [{\citenamefont {Zaletel}\ \emph {et~al.}(2013)\citenamefont {Zaletel}, \citenamefont {Mong},\ and\ \citenamefont {Pollmann}}]{ZaletelMongPollmann2013}%
  \BibitemOpen
  \bibfield  {author} {\bibinfo {author} {\bibfnamefont {M.~P.}\ \bibnamefont {Zaletel}}, \bibinfo {author} {\bibfnamefont {R.~S.~K.}\ \bibnamefont {Mong}},\ and\ \bibinfo {author} {\bibfnamefont {F.}~\bibnamefont {Pollmann}},\ }\bibfield  {title} {\bibinfo {title} {Topological characterization of fractional quantum hall ground states from microscopic hamiltonians},\ }\href {https://doi.org/10.1103/PhysRevLett.110.236801} {\bibfield  {journal} {\bibinfo  {journal} {Phys. Rev. Lett.}\ }\textbf {\bibinfo {volume} {110}},\ \bibinfo {pages} {236801} (\bibinfo {year} {2013})}\BibitemShut {NoStop}%
\bibitem [{\citenamefont {Trugman}\ and\ \citenamefont {Kivelson}(1985)}]{TrugmanKivelson1985}%
  \BibitemOpen
  \bibfield  {author} {\bibinfo {author} {\bibfnamefont {S.~A.}\ \bibnamefont {Trugman}}\ and\ \bibinfo {author} {\bibfnamefont {S.}~\bibnamefont {Kivelson}},\ }\bibfield  {title} {\bibinfo {title} {Exact results for the fractional quantum hall effect with general interactions},\ }\href {https://doi.org/10.1103/PhysRevB.31.5280} {\bibfield  {journal} {\bibinfo  {journal} {Phys. Rev. B}\ }\textbf {\bibinfo {volume} {31}},\ \bibinfo {pages} {5280} (\bibinfo {year} {1985})}\BibitemShut {NoStop}%
\bibitem [{\citenamefont {Rezayi}\ and\ \citenamefont {Haldane}(2000)}]{RezayiHaldane2000}%
  \BibitemOpen
  \bibfield  {author} {\bibinfo {author} {\bibfnamefont {E.~H.}\ \bibnamefont {Rezayi}}\ and\ \bibinfo {author} {\bibfnamefont {F.~D.~M.}\ \bibnamefont {Haldane}},\ }\bibfield  {title} {\bibinfo {title} {Incompressible paired hall state, stripe order, and the composite fermion liquid phase in half-filled landau levels},\ }\href {https://doi.org/10.1103/PhysRevLett.84.4685} {\bibfield  {journal} {\bibinfo  {journal} {Phys. Rev. Lett.}\ }\textbf {\bibinfo {volume} {84}},\ \bibinfo {pages} {4685} (\bibinfo {year} {2000})},\ \Eprint {https://arxiv.org/abs/cond-mat/9906137} {arXiv:cond-mat/9906137} \BibitemShut {NoStop}%
\bibitem [{\citenamefont {Laturia}\ \emph {et~al.}(2018)\citenamefont {Laturia}, \citenamefont {Van~de Put},\ and\ \citenamefont {Vandenberghe}}]{LaturiaEtAl2018}%
  \BibitemOpen
  \bibfield  {author} {\bibinfo {author} {\bibfnamefont {A.}~\bibnamefont {Laturia}}, \bibinfo {author} {\bibfnamefont {M.~L.}\ \bibnamefont {Van~de Put}},\ and\ \bibinfo {author} {\bibfnamefont {W.~G.}\ \bibnamefont {Vandenberghe}},\ }\bibfield  {title} {\bibinfo {title} {Dielectric properties of hexagonal boron nitride and transition metal dichalcogenides: from monolayer to bulk},\ }\href {https://doi.org/10.1038/s41699-018-0050-x} {\bibfield  {journal} {\bibinfo  {journal} {npj 2D Materials and Applications}\ }\textbf {\bibinfo {volume} {2}},\ \bibinfo {pages} {6} (\bibinfo {year} {2018})}\BibitemShut {NoStop}%
\bibitem [{\citenamefont {Nakamura}\ \emph {et~al.}(2019)\citenamefont {Nakamura}, \citenamefont {Fallahi}, \citenamefont {Sahasrabudhe}, \citenamefont {Rahman}, \citenamefont {Liang}, \citenamefont {Gardner},\ and\ \citenamefont {Manfra}}]{NakamuraEtAl2019AB}%
  \BibitemOpen
  \bibfield  {author} {\bibinfo {author} {\bibfnamefont {J.}~\bibnamefont {Nakamura}}, \bibinfo {author} {\bibfnamefont {S.}~\bibnamefont {Fallahi}}, \bibinfo {author} {\bibfnamefont {H.}~\bibnamefont {Sahasrabudhe}}, \bibinfo {author} {\bibfnamefont {R.}~\bibnamefont {Rahman}}, \bibinfo {author} {\bibfnamefont {S.}~\bibnamefont {Liang}}, \bibinfo {author} {\bibfnamefont {G.~C.}\ \bibnamefont {Gardner}},\ and\ \bibinfo {author} {\bibfnamefont {M.~J.}\ \bibnamefont {Manfra}},\ }\bibfield  {title} {\bibinfo {title} {Aharonov--bohm interference of fractional quantum hall edge modes},\ }\href {https://doi.org/10.1038/s41567-019-0441-8} {\bibfield  {journal} {\bibinfo  {journal} {Nature Physics}\ }\textbf {\bibinfo {volume} {15}},\ \bibinfo {pages} {563} (\bibinfo {year} {2019})}\BibitemShut {NoStop}%
\bibitem [{\citenamefont {Cohen}\ \emph {et~al.}(2023)\citenamefont {Cohen}, \citenamefont {Samuelson}, \citenamefont {Wang}, \citenamefont {Klocke}, \citenamefont {Reeves}, \citenamefont {Taniguchi}, \citenamefont {Watanabe}, \citenamefont {Vijay}, \citenamefont {Zaletel},\ and\ \citenamefont {Young}}]{CohenEtAl2023}%
  \BibitemOpen
  \bibfield  {author} {\bibinfo {author} {\bibfnamefont {L.~A.}\ \bibnamefont {Cohen}}, \bibinfo {author} {\bibfnamefont {N.~L.}\ \bibnamefont {Samuelson}}, \bibinfo {author} {\bibfnamefont {T.}~\bibnamefont {Wang}}, \bibinfo {author} {\bibfnamefont {K.}~\bibnamefont {Klocke}}, \bibinfo {author} {\bibfnamefont {C.~C.}\ \bibnamefont {Reeves}}, \bibinfo {author} {\bibfnamefont {T.}~\bibnamefont {Taniguchi}}, \bibinfo {author} {\bibfnamefont {K.}~\bibnamefont {Watanabe}}, \bibinfo {author} {\bibfnamefont {S.}~\bibnamefont {Vijay}}, \bibinfo {author} {\bibfnamefont {M.~P.}\ \bibnamefont {Zaletel}},\ and\ \bibinfo {author} {\bibfnamefont {A.~F.}\ \bibnamefont {Young}},\ }\bibfield  {title} {\bibinfo {title} {Nanoscale electrostatic control in ultraclean van der waals heterostructures by local anodic oxidation of graphite gates},\ }\href {https://doi.org/10.1038/s41567-023-02114-3} {\bibfield  {journal} {\bibinfo  {journal} {Nature Physics}\ }\textbf {\bibinfo {volume} {19}},\ \bibinfo {pages} {1502} (\bibinfo
  {year} {2023})}\BibitemShut {NoStop}%
\bibitem [{\citenamefont {Werkmeister}\ \emph {et~al.}(2024)\citenamefont {Werkmeister}, \citenamefont {Ehrets}, \citenamefont {Ronen}, \citenamefont {Wesson}, \citenamefont {Najafabadi}, \citenamefont {Wei}, \citenamefont {Watanabe}, \citenamefont {Taniguchi}, \citenamefont {Feldman}, \citenamefont {Halperin}, \citenamefont {Yacoby},\ and\ \citenamefont {Kim}}]{WerkmeisterEtAl2024}%
  \BibitemOpen
  \bibfield  {author} {\bibinfo {author} {\bibfnamefont {T.}~\bibnamefont {Werkmeister}}, \bibinfo {author} {\bibfnamefont {J.~R.}\ \bibnamefont {Ehrets}}, \bibinfo {author} {\bibfnamefont {Y.}~\bibnamefont {Ronen}}, \bibinfo {author} {\bibfnamefont {M.~E.}\ \bibnamefont {Wesson}}, \bibinfo {author} {\bibfnamefont {D.}~\bibnamefont {Najafabadi}}, \bibinfo {author} {\bibfnamefont {Z.}~\bibnamefont {Wei}}, \bibinfo {author} {\bibfnamefont {K.}~\bibnamefont {Watanabe}}, \bibinfo {author} {\bibfnamefont {T.}~\bibnamefont {Taniguchi}}, \bibinfo {author} {\bibfnamefont {D.~E.}\ \bibnamefont {Feldman}}, \bibinfo {author} {\bibfnamefont {B.~I.}\ \bibnamefont {Halperin}}, \bibinfo {author} {\bibfnamefont {A.}~\bibnamefont {Yacoby}},\ and\ \bibinfo {author} {\bibfnamefont {P.}~\bibnamefont {Kim}},\ }\bibfield  {title} {\bibinfo {title} {Strongly coupled edge states in a graphene quantum hall interferometer},\ }\href {https://doi.org/10.1038/s41467-024-50695-1} {\bibfield  {journal} {\bibinfo  {journal} {Nature
  Communications}\ }\textbf {\bibinfo {volume} {15}},\ \bibinfo {pages} {6533} (\bibinfo {year} {2024})}\BibitemShut {NoStop}%
\bibitem [{\citenamefont {Zeng}\ \emph {et~al.}(2023)\citenamefont {Zeng}, \citenamefont {Xia}, \citenamefont {Kang}, \citenamefont {Zhu}, \citenamefont {Kn\"uppel}, \citenamefont {Vaswani}, \citenamefont {Watanabe}, \citenamefont {Taniguchi}, \citenamefont {Mak},\ and\ \citenamefont {Shan}}]{ZengEtAl2023MoTe2}%
  \BibitemOpen
  \bibfield  {author} {\bibinfo {author} {\bibfnamefont {Y.}~\bibnamefont {Zeng}}, \bibinfo {author} {\bibfnamefont {Z.}~\bibnamefont {Xia}}, \bibinfo {author} {\bibfnamefont {K.}~\bibnamefont {Kang}}, \bibinfo {author} {\bibfnamefont {J.}~\bibnamefont {Zhu}}, \bibinfo {author} {\bibfnamefont {P.}~\bibnamefont {Kn\"uppel}}, \bibinfo {author} {\bibfnamefont {C.}~\bibnamefont {Vaswani}}, \bibinfo {author} {\bibfnamefont {K.}~\bibnamefont {Watanabe}}, \bibinfo {author} {\bibfnamefont {T.}~\bibnamefont {Taniguchi}}, \bibinfo {author} {\bibfnamefont {K.~F.}\ \bibnamefont {Mak}},\ and\ \bibinfo {author} {\bibfnamefont {J.}~\bibnamefont {Shan}},\ }\bibfield  {title} {\bibinfo {title} {Thermodynamic evidence of fractional chern insulator in moir\'e {MoTe}$_2$},\ }\href {https://doi.org/10.1038/s41586-023-06452-3} {\bibfield  {journal} {\bibinfo  {journal} {Nature}\ }\textbf {\bibinfo {volume} {622}},\ \bibinfo {pages} {69} (\bibinfo {year} {2023})}\BibitemShut {NoStop}%
\bibitem [{\citenamefont {Aronson}\ \emph {et~al.}(2025)\citenamefont {Aronson}, \citenamefont {Han}, \citenamefont {Lu}, \citenamefont {Yao}, \citenamefont {Butler}, \citenamefont {Watanabe}, \citenamefont {Taniguchi}, \citenamefont {Ju},\ and\ \citenamefont {Ashoori}}]{AronsonEtAl2025}%
  \BibitemOpen
  \bibfield  {author} {\bibinfo {author} {\bibfnamefont {S.~H.}\ \bibnamefont {Aronson}}, \bibinfo {author} {\bibfnamefont {T.}~\bibnamefont {Han}}, \bibinfo {author} {\bibfnamefont {Z.}~\bibnamefont {Lu}}, \bibinfo {author} {\bibfnamefont {Y.}~\bibnamefont {Yao}}, \bibinfo {author} {\bibfnamefont {J.~P.}\ \bibnamefont {Butler}}, \bibinfo {author} {\bibfnamefont {K.}~\bibnamefont {Watanabe}}, \bibinfo {author} {\bibfnamefont {T.}~\bibnamefont {Taniguchi}}, \bibinfo {author} {\bibfnamefont {L.}~\bibnamefont {Ju}},\ and\ \bibinfo {author} {\bibfnamefont {R.~C.}\ \bibnamefont {Ashoori}},\ }\bibfield  {title} {\bibinfo {title} {Displacement field-controlled fractional chern insulators and charge density waves in a graphene/hbn moir\'e superlattice},\ }\href {https://doi.org/10.1103/75gl-jzl6} {\bibfield  {journal} {\bibinfo  {journal} {Phys. Rev. X}\ }\textbf {\bibinfo {volume} {15}},\ \bibinfo {pages} {031026} (\bibinfo {year} {2025})}\BibitemShut {NoStop}%
\bibitem [{\citenamefont {Waters}\ \emph {et~al.}(2025)\citenamefont {Waters}, \citenamefont {Okounkova}, \citenamefont {Su}, \citenamefont {Zhou}, \citenamefont {Yao}, \citenamefont {Watanabe}, \citenamefont {Taniguchi}, \citenamefont {Xu}, \citenamefont {Zhang}, \citenamefont {Folk},\ and\ \citenamefont {Yankowitz}}]{WatersEtAl2025Pentalayer}%
  \BibitemOpen
  \bibfield  {author} {\bibinfo {author} {\bibfnamefont {D.}~\bibnamefont {Waters}}, \bibinfo {author} {\bibfnamefont {A.}~\bibnamefont {Okounkova}}, \bibinfo {author} {\bibfnamefont {R.}~\bibnamefont {Su}}, \bibinfo {author} {\bibfnamefont {B.}~\bibnamefont {Zhou}}, \bibinfo {author} {\bibfnamefont {J.}~\bibnamefont {Yao}}, \bibinfo {author} {\bibfnamefont {K.}~\bibnamefont {Watanabe}}, \bibinfo {author} {\bibfnamefont {T.}~\bibnamefont {Taniguchi}}, \bibinfo {author} {\bibfnamefont {X.}~\bibnamefont {Xu}}, \bibinfo {author} {\bibfnamefont {Y.-H.}\ \bibnamefont {Zhang}}, \bibinfo {author} {\bibfnamefont {J.}~\bibnamefont {Folk}},\ and\ \bibinfo {author} {\bibfnamefont {M.}~\bibnamefont {Yankowitz}},\ }\bibfield  {title} {\bibinfo {title} {Chern insulators at integer and fractional filling in moir\'e pentalayer graphene},\ }\href {https://doi.org/10.1103/PhysRevX.15.011045} {\bibfield  {journal} {\bibinfo  {journal} {Phys. Rev. X}\ }\textbf {\bibinfo {volume} {15}},\ \bibinfo {pages} {011045} (\bibinfo {year}
  {2025})}\BibitemShut {NoStop}%
\bibitem [{\citenamefont {Lu}\ \emph {et~al.}(2024)\citenamefont {Lu}, \citenamefont {Han}, \citenamefont {Yao}, \citenamefont {Reddy}, \citenamefont {Yang}, \citenamefont {Seo}, \citenamefont {Watanabe}, \citenamefont {Taniguchi}, \citenamefont {Fu},\ and\ \citenamefont {Ju}}]{LuEtAl2024FQAH}%
  \BibitemOpen
  \bibfield  {author} {\bibinfo {author} {\bibfnamefont {Z.}~\bibnamefont {Lu}}, \bibinfo {author} {\bibfnamefont {T.}~\bibnamefont {Han}}, \bibinfo {author} {\bibfnamefont {Y.}~\bibnamefont {Yao}}, \bibinfo {author} {\bibfnamefont {A.~P.}\ \bibnamefont {Reddy}}, \bibinfo {author} {\bibfnamefont {J.}~\bibnamefont {Yang}}, \bibinfo {author} {\bibfnamefont {J.}~\bibnamefont {Seo}}, \bibinfo {author} {\bibfnamefont {K.}~\bibnamefont {Watanabe}}, \bibinfo {author} {\bibfnamefont {T.}~\bibnamefont {Taniguchi}}, \bibinfo {author} {\bibfnamefont {L.}~\bibnamefont {Fu}},\ and\ \bibinfo {author} {\bibfnamefont {L.}~\bibnamefont {Ju}},\ }\bibfield  {title} {\bibinfo {title} {Fractional quantum anomalous hall effect in multilayer graphene},\ }\href {https://doi.org/10.1038/s41586-023-07010-7} {\bibfield  {journal} {\bibinfo  {journal} {Nature}\ }\textbf {\bibinfo {volume} {626}},\ \bibinfo {pages} {759} (\bibinfo {year} {2024})}\BibitemShut {NoStop}%
\bibitem [{Note2()}]{Note2}%
  \BibitemOpen
  \bibinfo {note} {Fractional exclusion statistics and residual interactions between anyons are not relevant in the dilute limit we consider.}\BibitemShut {Stop}%
\bibitem [{\citenamefont {Polyakov}\ and\ \citenamefont {Shklovskii}(1994)}]{PolyakovShklovskii1994}%
  \BibitemOpen
  \bibfield  {author} {\bibinfo {author} {\bibfnamefont {D.~G.}\ \bibnamefont {Polyakov}}\ and\ \bibinfo {author} {\bibfnamefont {B.~I.}\ \bibnamefont {Shklovskii}},\ }\bibfield  {title} {\bibinfo {title} {Activated conductivity in the quantum hall effect},\ }\href {https://doi.org/10.1103/PhysRevLett.73.1150} {\bibfield  {journal} {\bibinfo  {journal} {Phys. Rev. Lett.}\ }\textbf {\bibinfo {volume} {73}},\ \bibinfo {pages} {1150} (\bibinfo {year} {1994})}\BibitemShut {NoStop}%
\bibitem [{\citenamefont {Saha}(1920)}]{Saha1920Ionization}%
  \BibitemOpen
  \bibfield  {author} {\bibinfo {author} {\bibfnamefont {M.~N.}\ \bibnamefont {Saha}},\ }\bibfield  {title} {\bibinfo {title} {Liii. ionization in the solar chromosphere},\ }\href {https://doi.org/10.1080/14786441008636148} {\bibfield  {journal} {\bibinfo  {journal} {The London, Edinburgh, and Dublin Philosophical Magazine and Journal of Science}\ }\bibinfo {series} {6},\ \textbf {\bibinfo {volume} {40}},\ \bibinfo {pages} {472} (\bibinfo {year} {1920})}\BibitemShut {NoStop}%
\bibitem [{\citenamefont {Eisenstein}\ \emph {et~al.}(1992)\citenamefont {Eisenstein}, \citenamefont {Pfeiffer},\ and\ \citenamefont {West}}]{EisensteinEtAl1992NegativeCompressibility}%
  \BibitemOpen
  \bibfield  {author} {\bibinfo {author} {\bibfnamefont {J.~P.}\ \bibnamefont {Eisenstein}}, \bibinfo {author} {\bibfnamefont {L.~N.}\ \bibnamefont {Pfeiffer}},\ and\ \bibinfo {author} {\bibfnamefont {K.~W.}\ \bibnamefont {West}},\ }\bibfield  {title} {\bibinfo {title} {Negative compressibility of interacting two-dimensional electron and quasiparticle gases},\ }\href {https://doi.org/10.1103/PhysRevLett.68.674} {\bibfield  {journal} {\bibinfo  {journal} {Physical Review Letters}\ }\textbf {\bibinfo {volume} {68}},\ \bibinfo {pages} {674} (\bibinfo {year} {1992})}\BibitemShut {NoStop}%
\bibitem [{\citenamefont {Assouline}\ \emph {et~al.}(2026)\citenamefont {Assouline}, \citenamefont {Wang}, \citenamefont {Yoo}, \citenamefont {Fan}, \citenamefont {Yang}, \citenamefont {Zhang}, \citenamefont {Taniguchi}, \citenamefont {Watanabe}, \citenamefont {Zaletel},\ and\ \citenamefont {Young}}]{AssoulineEtAl2026Entropy}%
  \BibitemOpen
  \bibfield  {author} {\bibinfo {author} {\bibfnamefont {A.}~\bibnamefont {Assouline}}, \bibinfo {author} {\bibfnamefont {T.}~\bibnamefont {Wang}}, \bibinfo {author} {\bibfnamefont {H.~M.}\ \bibnamefont {Yoo}}, \bibinfo {author} {\bibfnamefont {R.}~\bibnamefont {Fan}}, \bibinfo {author} {\bibfnamefont {F.}~\bibnamefont {Yang}}, \bibinfo {author} {\bibfnamefont {R.}~\bibnamefont {Zhang}}, \bibinfo {author} {\bibfnamefont {T.}~\bibnamefont {Taniguchi}}, \bibinfo {author} {\bibfnamefont {K.}~\bibnamefont {Watanabe}}, \bibinfo {author} {\bibfnamefont {M.~P.}\ \bibnamefont {Zaletel}},\ and\ \bibinfo {author} {\bibfnamefont {A.~F.}\ \bibnamefont {Young}},\ }\bibfield  {title} {\bibinfo {title} {Entropy of strongly correlated electrons in a partially filled landau level},\ }\bibfield  {journal} {\bibinfo  {journal} {Physical Review X}\ }\href {https://doi.org/10.1103/jpf1-4zdk} {10.1103/jpf1-4zdk} (\bibinfo {year} {2026}),\ \bibinfo {note} {in press}\BibitemShut {NoStop}%
\bibitem [{\citenamefont {Hu}\ \emph {et~al.}(2025)\citenamefont {Hu}, \citenamefont {Tsui}, \citenamefont {He}, \citenamefont {Kamber}, \citenamefont {Wang}, \citenamefont {Mohammadi}, \citenamefont {Watanabe}, \citenamefont {Taniguchi}, \citenamefont {Papi\'c}, \citenamefont {Zaletel},\ and\ \citenamefont {Yazdani}}]{HuEtAl2025}%
  \BibitemOpen
  \bibfield  {author} {\bibinfo {author} {\bibfnamefont {Y.}~\bibnamefont {Hu}}, \bibinfo {author} {\bibfnamefont {Y.-C.}\ \bibnamefont {Tsui}}, \bibinfo {author} {\bibfnamefont {M.}~\bibnamefont {He}}, \bibinfo {author} {\bibfnamefont {U.}~\bibnamefont {Kamber}}, \bibinfo {author} {\bibfnamefont {T.}~\bibnamefont {Wang}}, \bibinfo {author} {\bibfnamefont {A.~S.}\ \bibnamefont {Mohammadi}}, \bibinfo {author} {\bibfnamefont {K.}~\bibnamefont {Watanabe}}, \bibinfo {author} {\bibfnamefont {T.}~\bibnamefont {Taniguchi}}, \bibinfo {author} {\bibfnamefont {Z.}~\bibnamefont {Papi\'c}}, \bibinfo {author} {\bibfnamefont {M.~P.}\ \bibnamefont {Zaletel}},\ and\ \bibinfo {author} {\bibfnamefont {A.}~\bibnamefont {Yazdani}},\ }\bibfield  {title} {\bibinfo {title} {High-resolution tunnelling spectroscopy of fractional quantum hall states},\ }\href {https://doi.org/10.1038/s41567-025-02830-y} {\bibfield  {journal} {\bibinfo  {journal} {Nature Physics}\ }\textbf {\bibinfo {volume} {21}},\ \bibinfo {pages} {716} (\bibinfo
  {year} {2025})}\BibitemShut {NoStop}%
\bibitem [{\citenamefont {Chiu}\ \emph {et~al.}(2025)\citenamefont {Chiu}, \citenamefont {Wang}, \citenamefont {Fan}, \citenamefont {Watanabe}, \citenamefont {Taniguchi}, \citenamefont {Liu}, \citenamefont {Zaletel},\ and\ \citenamefont {Yazdani}}]{ChiuEtAl2025}%
  \BibitemOpen
  \bibfield  {author} {\bibinfo {author} {\bibfnamefont {C.-L.}\ \bibnamefont {Chiu}}, \bibinfo {author} {\bibfnamefont {T.}~\bibnamefont {Wang}}, \bibinfo {author} {\bibfnamefont {R.}~\bibnamefont {Fan}}, \bibinfo {author} {\bibfnamefont {K.}~\bibnamefont {Watanabe}}, \bibinfo {author} {\bibfnamefont {T.}~\bibnamefont {Taniguchi}}, \bibinfo {author} {\bibfnamefont {X.}~\bibnamefont {Liu}}, \bibinfo {author} {\bibfnamefont {M.~P.}\ \bibnamefont {Zaletel}},\ and\ \bibinfo {author} {\bibfnamefont {A.}~\bibnamefont {Yazdani}},\ }\bibfield  {title} {\bibinfo {title} {High spatial resolution charge sensing of quantum hall states},\ }\href {https://doi.org/10.1073/pnas.2424781122} {\bibfield  {journal} {\bibinfo  {journal} {Proceedings of the National Academy of Sciences of the United States of America}\ }\textbf {\bibinfo {volume} {122}},\ \bibinfo {pages} {e2424781122} (\bibinfo {year} {2025})}\BibitemShut {NoStop}%
\bibitem [{\citenamefont {Assouline}\ \emph {et~al.}(2024)\citenamefont {Assouline}, \citenamefont {Wang}, \citenamefont {Zhou}, \citenamefont {Cohen}, \citenamefont {Yang}, \citenamefont {Zhang}, \citenamefont {Taniguchi}, \citenamefont {Watanabe}, \citenamefont {Mong}, \citenamefont {Zaletel},\ and\ \citenamefont {Young}}]{AssoulineEtAl2024}%
  \BibitemOpen
  \bibfield  {author} {\bibinfo {author} {\bibfnamefont {A.}~\bibnamefont {Assouline}}, \bibinfo {author} {\bibfnamefont {T.}~\bibnamefont {Wang}}, \bibinfo {author} {\bibfnamefont {H.}~\bibnamefont {Zhou}}, \bibinfo {author} {\bibfnamefont {L.~A.}\ \bibnamefont {Cohen}}, \bibinfo {author} {\bibfnamefont {F.}~\bibnamefont {Yang}}, \bibinfo {author} {\bibfnamefont {R.}~\bibnamefont {Zhang}}, \bibinfo {author} {\bibfnamefont {T.}~\bibnamefont {Taniguchi}}, \bibinfo {author} {\bibfnamefont {K.}~\bibnamefont {Watanabe}}, \bibinfo {author} {\bibfnamefont {R.~S.~K.}\ \bibnamefont {Mong}}, \bibinfo {author} {\bibfnamefont {M.~P.}\ \bibnamefont {Zaletel}},\ and\ \bibinfo {author} {\bibfnamefont {A.~F.}\ \bibnamefont {Young}},\ }\bibfield  {title} {\bibinfo {title} {Energy gap of the even-denominator fractional quantum hall state in bilayer graphene},\ }\href {https://doi.org/10.1103/PhysRevLett.132.046603} {\bibfield  {journal} {\bibinfo  {journal} {Phys. Rev. Lett.}\ }\textbf {\bibinfo {volume} {132}},\ \bibinfo
  {pages} {046603} (\bibinfo {year} {2024})}\BibitemShut {NoStop}%
\bibitem [{\citenamefont {Cooper}\ and\ \citenamefont {Stern}(2009)}]{Cooper2009}%
  \BibitemOpen
  \bibfield  {author} {\bibinfo {author} {\bibfnamefont {N.}~\bibnamefont {Cooper}}\ and\ \bibinfo {author} {\bibfnamefont {A.}~\bibnamefont {Stern}},\ }\bibfield  {title} {\bibinfo {title} {Observable bulk signatures of non-abelian quantum hall states},\ }\href {https://doi.org/10.1103/PhysRevLett.102.176807} {\bibfield  {journal} {\bibinfo  {journal} {Physical Review Letters}\ }\textbf {\bibinfo {volume} {102}},\ \bibinfo {pages} {176807} (\bibinfo {year} {2009})}\BibitemShut {NoStop}%
\bibitem [{\citenamefont {Yang}\ and\ \citenamefont {Halperin}(2009)}]{YangHalperin2009Thermopower}%
  \BibitemOpen
  \bibfield  {author} {\bibinfo {author} {\bibfnamefont {K.}~\bibnamefont {Yang}}\ and\ \bibinfo {author} {\bibfnamefont {B.~I.}\ \bibnamefont {Halperin}},\ }\bibfield  {title} {\bibinfo {title} {Thermopower as a possible probe of non-abelian quasiparticle statistics in fractional quantum hall liquids},\ }\href {https://doi.org/10.1103/PhysRevB.79.115317} {\bibfield  {journal} {\bibinfo  {journal} {Physical Review B}\ }\textbf {\bibinfo {volume} {79}},\ \bibinfo {pages} {115317} (\bibinfo {year} {2009})},\ \Eprint {https://arxiv.org/abs/0901.1429} {arXiv:0901.1429 [cond-mat.mes-hall]} \BibitemShut {NoStop}%
\bibitem [{Note3()}]{Note3}%
  \BibitemOpen
  \bibinfo {note} {Estimated from $C \simeq 2\epsilon _z A/d_g$ using $d_g=\SI {40}{nm}$, $\epsilon _z=3$, $\epsilon _{xy}=6.6$, $B=\SI {14}{T}$, and $E_{\protect \rm bind}=0.004E_C$.}\BibitemShut {Stop}%
\bibitem [{\citenamefont {Luca}\ \emph {et~al.}(2025)\citenamefont {Luca}, \citenamefont {Hajigeorgiou}, \citenamefont {Zhou}, \citenamefont {Lotri{\v c}}, \citenamefont {Feng}, \citenamefont {Watanabe}, \citenamefont {Taniguchi}, \citenamefont {Simon},\ and\ \citenamefont {Banerjee}}]{DiLucaEtAl2025FractionalCoulombmeter}%
  \BibitemOpen
  \bibfield  {author} {\bibinfo {author} {\bibfnamefont {M.~D.}\ \bibnamefont {Luca}}, \bibinfo {author} {\bibfnamefont {E.}~\bibnamefont {Hajigeorgiou}}, \bibinfo {author} {\bibfnamefont {Z.}~\bibnamefont {Zhou}}, \bibinfo {author} {\bibfnamefont {T.}~\bibnamefont {Lotri{\v c}}}, \bibinfo {author} {\bibfnamefont {T.}~\bibnamefont {Feng}}, \bibinfo {author} {\bibfnamefont {K.}~\bibnamefont {Watanabe}}, \bibinfo {author} {\bibfnamefont {T.}~\bibnamefont {Taniguchi}}, \bibinfo {author} {\bibfnamefont {S.~H.}\ \bibnamefont {Simon}},\ and\ \bibinfo {author} {\bibfnamefont {M.}~\bibnamefont {Banerjee}},\ }\href@noop {} {\bibinfo {title} {Quantum hall antidot as a fractional coulombmeter}} (\bibinfo {year} {2025}),\ \Eprint {https://arxiv.org/abs/2509.04209} {arXiv:2509.04209 [cond-mat.mes-hall]} \BibitemShut {NoStop}%
\bibitem [{\citenamefont {Halperin}\ \emph {et~al.}(2011)\citenamefont {Halperin}, \citenamefont {Stern}, \citenamefont {Neder},\ and\ \citenamefont {Rosenow}}]{HalperinEtAl2011}%
  \BibitemOpen
  \bibfield  {author} {\bibinfo {author} {\bibfnamefont {B.~I.}\ \bibnamefont {Halperin}}, \bibinfo {author} {\bibfnamefont {A.}~\bibnamefont {Stern}}, \bibinfo {author} {\bibfnamefont {I.}~\bibnamefont {Neder}},\ and\ \bibinfo {author} {\bibfnamefont {B.}~\bibnamefont {Rosenow}},\ }\bibfield  {title} {\bibinfo {title} {Theory of the fabry-p\'erot quantum hall interferometer},\ }\href {https://doi.org/10.1103/PhysRevB.83.155440} {\bibfield  {journal} {\bibinfo  {journal} {Phys. Rev. B}\ }\textbf {\bibinfo {volume} {83}},\ \bibinfo {pages} {155440} (\bibinfo {year} {2011})}\BibitemShut {NoStop}%
\bibitem [{\citenamefont {Rosenow}\ and\ \citenamefont {Halperin}(2007)}]{RosenowHalperin2007}%
  \BibitemOpen
  \bibfield  {author} {\bibinfo {author} {\bibfnamefont {B.}~\bibnamefont {Rosenow}}\ and\ \bibinfo {author} {\bibfnamefont {B.~I.}\ \bibnamefont {Halperin}},\ }\bibfield  {title} {\bibinfo {title} {Influence of interactions on flux and back-gate period of quantum hall interferometers},\ }\href {https://doi.org/10.1103/PhysRevLett.98.106801} {\bibfield  {journal} {\bibinfo  {journal} {Phys. Rev. Lett.}\ }\textbf {\bibinfo {volume} {98}},\ \bibinfo {pages} {106801} (\bibinfo {year} {2007})}\BibitemShut {NoStop}%
\bibitem [{\citenamefont {Nakamura}\ \emph {et~al.}(2022)\citenamefont {Nakamura}, \citenamefont {Liang}, \citenamefont {Gardner},\ and\ \citenamefont {Manfra}}]{NakamuraEtAl2022}%
  \BibitemOpen
  \bibfield  {author} {\bibinfo {author} {\bibfnamefont {J.}~\bibnamefont {Nakamura}}, \bibinfo {author} {\bibfnamefont {S.}~\bibnamefont {Liang}}, \bibinfo {author} {\bibfnamefont {G.~C.}\ \bibnamefont {Gardner}},\ and\ \bibinfo {author} {\bibfnamefont {M.~J.}\ \bibnamefont {Manfra}},\ }\bibfield  {title} {\bibinfo {title} {Impact of bulk-edge coupling on observation of anyonic braiding statistics in quantum hall interferometers},\ }\href {https://doi.org/10.1038/s41467-022-27958-w} {\bibfield  {journal} {\bibinfo  {journal} {Nature Communications}\ }\textbf {\bibinfo {volume} {13}},\ \bibinfo {pages} {344} (\bibinfo {year} {2022})}\BibitemShut {NoStop}%
\bibitem [{\citenamefont {Samuelson}\ \emph {et~al.}(2026)\citenamefont {Samuelson}, \citenamefont {Cohen}, \citenamefont {Wang}, \citenamefont {Blanch}, \citenamefont {Taniguchi}, \citenamefont {Watanabe}, \citenamefont {Zaletel},\ and\ \citenamefont {Young}}]{SamuelsonEtAl2026Slow}%
  \BibitemOpen
  \bibfield  {author} {\bibinfo {author} {\bibfnamefont {N.~L.}\ \bibnamefont {Samuelson}}, \bibinfo {author} {\bibfnamefont {L.~A.}\ \bibnamefont {Cohen}}, \bibinfo {author} {\bibfnamefont {W.}~\bibnamefont {Wang}}, \bibinfo {author} {\bibfnamefont {S.}~\bibnamefont {Blanch}}, \bibinfo {author} {\bibfnamefont {T.}~\bibnamefont {Taniguchi}}, \bibinfo {author} {\bibfnamefont {K.}~\bibnamefont {Watanabe}}, \bibinfo {author} {\bibfnamefont {M.~P.}\ \bibnamefont {Zaletel}},\ and\ \bibinfo {author} {\bibfnamefont {A.~F.}\ \bibnamefont {Young}},\ }\bibfield  {title} {\bibinfo {title} {Slow quasiparticle dynamics and anyonic statistics in a fractional quantum hall fabry-p\'erot interferometer},\ }\href {https://doi.org/10.1103/fwjg-mx9h} {\bibfield  {journal} {\bibinfo  {journal} {Phys. Rev. X}\ }\textbf {\bibinfo {volume} {16}},\ \bibinfo {pages} {011062} (\bibinfo {year} {2026})}\BibitemShut {NoStop}%
\bibitem [{\citenamefont {Henzinger}\ \emph {et~al.}(2026)\citenamefont {Henzinger}, \citenamefont {Ehrets}, \citenamefont {Fushio}, \citenamefont {Dong}, \citenamefont {Werkmeister}, \citenamefont {Wesson}, \citenamefont {Watanabe}, \citenamefont {Taniguchi}, \citenamefont {Vishwanath}, \citenamefont {Halperin}, \citenamefont {Yacoby},\ and\ \citenamefont {Kim}}]{HenzingerEtAl2026Control}%
  \BibitemOpen
  \bibfield  {author} {\bibinfo {author} {\bibfnamefont {C.~E.}\ \bibnamefont {Henzinger}}, \bibinfo {author} {\bibfnamefont {J.~R.}\ \bibnamefont {Ehrets}}, \bibinfo {author} {\bibfnamefont {R.}~\bibnamefont {Fushio}}, \bibinfo {author} {\bibfnamefont {J.}~\bibnamefont {Dong}}, \bibinfo {author} {\bibfnamefont {T.}~\bibnamefont {Werkmeister}}, \bibinfo {author} {\bibfnamefont {M.~E.}\ \bibnamefont {Wesson}}, \bibinfo {author} {\bibfnamefont {K.}~\bibnamefont {Watanabe}}, \bibinfo {author} {\bibfnamefont {T.}~\bibnamefont {Taniguchi}}, \bibinfo {author} {\bibfnamefont {A.}~\bibnamefont {Vishwanath}}, \bibinfo {author} {\bibfnamefont {B.~I.}\ \bibnamefont {Halperin}}, \bibinfo {author} {\bibfnamefont {A.}~\bibnamefont {Yacoby}},\ and\ \bibinfo {author} {\bibfnamefont {P.}~\bibnamefont {Kim}},\ }\href@noop {} {\bibinfo {title} {Controlled localization of anyons in a graphene quantum hall interferometer}} (\bibinfo {year} {2026}),\ \Eprint {https://arxiv.org/abs/2603.11182} {arXiv:2603.11182 [cond-mat.str-el]}
  \BibitemShut {NoStop}%
\bibitem [{\citenamefont {Bid}\ \emph {et~al.}(2009)\citenamefont {Bid}, \citenamefont {Ofek}, \citenamefont {Heiblum}, \citenamefont {Umansky},\ and\ \citenamefont {Mahalu}}]{BidEtAl2009}%
  \BibitemOpen
  \bibfield  {author} {\bibinfo {author} {\bibfnamefont {A.}~\bibnamefont {Bid}}, \bibinfo {author} {\bibfnamefont {N.}~\bibnamefont {Ofek}}, \bibinfo {author} {\bibfnamefont {M.}~\bibnamefont {Heiblum}}, \bibinfo {author} {\bibfnamefont {V.}~\bibnamefont {Umansky}},\ and\ \bibinfo {author} {\bibfnamefont {D.}~\bibnamefont {Mahalu}},\ }\bibfield  {title} {\bibinfo {title} {Shot noise and charge at the 2/3 composite fractional quantum hall state},\ }\href {https://doi.org/10.1103/PhysRevLett.103.236802} {\bibfield  {journal} {\bibinfo  {journal} {Phys. Rev. Lett.}\ }\textbf {\bibinfo {volume} {103}},\ \bibinfo {pages} {236802} (\bibinfo {year} {2009})}\BibitemShut {NoStop}%
\bibitem [{\citenamefont {Ghosh}\ \emph {et~al.}(2025)\citenamefont {Ghosh}, \citenamefont {Labendik}, \citenamefont {Umansky}, \citenamefont {Heiblum},\ and\ \citenamefont {Mross}}]{GhoshEtAl2025}%
  \BibitemOpen
  \bibfield  {author} {\bibinfo {author} {\bibfnamefont {B.}~\bibnamefont {Ghosh}}, \bibinfo {author} {\bibfnamefont {M.}~\bibnamefont {Labendik}}, \bibinfo {author} {\bibfnamefont {V.}~\bibnamefont {Umansky}}, \bibinfo {author} {\bibfnamefont {M.}~\bibnamefont {Heiblum}},\ and\ \bibinfo {author} {\bibfnamefont {D.~F.}\ \bibnamefont {Mross}},\ }\bibfield  {title} {\bibinfo {title} {Coherent bunching of anyons and dissociation in an interference experiment},\ }\href {https://doi.org/10.1038/s41586-025-09143-3} {\bibfield  {journal} {\bibinfo  {journal} {Nature}\ }\textbf {\bibinfo {volume} {642}},\ \bibinfo {pages} {922} (\bibinfo {year} {2025})}\BibitemShut {NoStop}%
\bibitem [{\citenamefont {De}\ \emph {et~al.}(2025)\citenamefont {De}, \citenamefont {Boudet}, \citenamefont {Nath}, \citenamefont {Kapfer}, \citenamefont {Farrer}, \citenamefont {Ritchie}, \citenamefont {Roulleau},\ and\ \citenamefont {Glattli}}]{DeEtAl2025}%
  \BibitemOpen
  \bibfield  {author} {\bibinfo {author} {\bibfnamefont {A.}~\bibnamefont {De}}, \bibinfo {author} {\bibfnamefont {C.}~\bibnamefont {Boudet}}, \bibinfo {author} {\bibfnamefont {J.}~\bibnamefont {Nath}}, \bibinfo {author} {\bibfnamefont {M.}~\bibnamefont {Kapfer}}, \bibinfo {author} {\bibfnamefont {I.}~\bibnamefont {Farrer}}, \bibinfo {author} {\bibfnamefont {D.~A.}\ \bibnamefont {Ritchie}}, \bibinfo {author} {\bibfnamefont {P.}~\bibnamefont {Roulleau}},\ and\ \bibinfo {author} {\bibfnamefont {D.~C.}\ \bibnamefont {Glattli}},\ }\bibfield  {title} {\bibinfo {title} {Electronic hong--ou--mandel interferences to unveil the 2/3 fractional quantum hall edge channel dynamics},\ }\href {https://doi.org/10.1038/s41467-025-63308-2} {\bibfield  {journal} {\bibinfo  {journal} {Nature Communications}\ }\textbf {\bibinfo {volume} {16}},\ \bibinfo {pages} {8466} (\bibinfo {year} {2025})}\BibitemShut {NoStop}%
\bibitem [{\citenamefont {Laughlin}(1988{\natexlab{a}})}]{Laughlin1988PRL}%
  \BibitemOpen
  \bibfield  {author} {\bibinfo {author} {\bibfnamefont {R.~B.}\ \bibnamefont {Laughlin}},\ }\bibfield  {title} {\bibinfo {title} {Superconducting ground state of noninteracting particles obeying fractional statistics},\ }\href {https://doi.org/10.1103/PhysRevLett.60.2677} {\bibfield  {journal} {\bibinfo  {journal} {Physical Review Letters}\ }\textbf {\bibinfo {volume} {60}},\ \bibinfo {pages} {2677} (\bibinfo {year} {1988}{\natexlab{a}})}\BibitemShut {NoStop}%
\bibitem [{\citenamefont {Laughlin}(1988{\natexlab{b}})}]{Laughlin1988Science}%
  \BibitemOpen
  \bibfield  {author} {\bibinfo {author} {\bibfnamefont {R.~B.}\ \bibnamefont {Laughlin}},\ }\bibfield  {title} {\bibinfo {title} {The relationship between high-temperature superconductivity and the fractional quantum hall effect},\ }\href {https://doi.org/10.1126/science.242.4878.525} {\bibfield  {journal} {\bibinfo  {journal} {Science}\ }\textbf {\bibinfo {volume} {242}},\ \bibinfo {pages} {525} (\bibinfo {year} {1988}{\natexlab{b}})}\BibitemShut {NoStop}%
\bibitem [{\citenamefont {Halperin}(1992)}]{HalperinAnyonSC1992}%
  \BibitemOpen
  \bibfield  {author} {\bibinfo {author} {\bibfnamefont {B.~I.}\ \bibnamefont {Halperin}},\ }\bibfield  {title} {\bibinfo {title} {Anyons in the quantized hall effect and in models of high temperature superconductors},\ }\href {https://doi.org/10.5169/seals-116400} {\bibfield  {journal} {\bibinfo  {journal} {Helvetica Physica Acta}\ }\textbf {\bibinfo {volume} {65}},\ \bibinfo {pages} {215} (\bibinfo {year} {1992})}\BibitemShut {NoStop}%
\bibitem [{\citenamefont {Lee}\ and\ \citenamefont {Fisher}(1991)}]{FisherLeeAnyonSC1991}%
  \BibitemOpen
  \bibfield  {author} {\bibinfo {author} {\bibfnamefont {D.-H.}\ \bibnamefont {Lee}}\ and\ \bibinfo {author} {\bibfnamefont {M.~P.~A.}\ \bibnamefont {Fisher}},\ }\bibfield  {title} {\bibinfo {title} {Anyon superconductivity and charge-vortex duality},\ }\href {https://doi.org/10.1142/S0217979291001061} {\bibfield  {journal} {\bibinfo  {journal} {Int. J. Mod. Phys. B}\ }\textbf {\bibinfo {volume} {5}},\ \bibinfo {pages} {2675} (\bibinfo {year} {1991})}\BibitemShut {NoStop}%
\bibitem [{\citenamefont {Shi}\ and\ \citenamefont {Senthil}(2025)}]{ShiSenthil2025}%
  \BibitemOpen
  \bibfield  {author} {\bibinfo {author} {\bibfnamefont {Z.~D.}\ \bibnamefont {Shi}}\ and\ \bibinfo {author} {\bibfnamefont {T.}~\bibnamefont {Senthil}},\ }\bibfield  {title} {\bibinfo {title} {Doping a fractional quantum anomalous hall insulator},\ }\href {https://doi.org/10.1103/kcm5-hx56} {\bibfield  {journal} {\bibinfo  {journal} {Phys. Rev. X}\ }\textbf {\bibinfo {volume} {15}},\ \bibinfo {pages} {031069} (\bibinfo {year} {2025})}\BibitemShut {NoStop}%
\bibitem [{\citenamefont {Pichler}\ \emph {et~al.}(2026)\citenamefont {Pichler}, \citenamefont {Kuhlenkamp}, \citenamefont {Knap},\ and\ \citenamefont {Vishwanath}}]{PichlerEtAl2025}%
  \BibitemOpen
  \bibfield  {author} {\bibinfo {author} {\bibfnamefont {F.}~\bibnamefont {Pichler}}, \bibinfo {author} {\bibfnamefont {C.}~\bibnamefont {Kuhlenkamp}}, \bibinfo {author} {\bibfnamefont {M.}~\bibnamefont {Knap}},\ and\ \bibinfo {author} {\bibfnamefont {A.}~\bibnamefont {Vishwanath}},\ }\bibfield  {title} {\bibinfo {title} {Microscopic mechanism of anyon superconductivity emerging from fractional chern insulators},\ }\href {https://doi.org/10.1016/j.newton.2025.100340} {\bibfield  {journal} {\bibinfo  {journal} {Newton}\ }\textbf {\bibinfo {volume} {2}},\ \bibinfo {pages} {100340} (\bibinfo {year} {2026})}\BibitemShut {NoStop}%
\bibitem [{\citenamefont {Wang}\ and\ \citenamefont {Zaletel}(2025)}]{WangZaletel2025}%
  \BibitemOpen
  \bibfield  {author} {\bibinfo {author} {\bibfnamefont {T.}~\bibnamefont {Wang}}\ and\ \bibinfo {author} {\bibfnamefont {M.~P.}\ \bibnamefont {Zaletel}},\ }\href {https://doi.org/10.48550/arXiv.2507.07921} {\bibinfo {title} {Chiral superconductivity near a fractional chern insulator}} (\bibinfo {year} {2025}),\ \Eprint {https://arxiv.org/abs/2507.07921} {arXiv:2507.07921 [cond-mat.str-el]} \BibitemShut {NoStop}%
\bibitem [{\citenamefont {Xu}\ \emph {et~al.}(2025)\citenamefont {Xu}, \citenamefont {Ji}, \citenamefont {Wang}, \citenamefont {Trung},\ and\ \citenamefont {Yang}}]{XuEtAl2025AnyonClusters}%
  \BibitemOpen
  \bibfield  {author} {\bibinfo {author} {\bibfnamefont {Q.}~\bibnamefont {Xu}}, \bibinfo {author} {\bibfnamefont {G.}~\bibnamefont {Ji}}, \bibinfo {author} {\bibfnamefont {Y.}~\bibnamefont {Wang}}, \bibinfo {author} {\bibfnamefont {H.~Q.}\ \bibnamefont {Trung}},\ and\ \bibinfo {author} {\bibfnamefont {B.}~\bibnamefont {Yang}},\ }\bibfield  {title} {\bibinfo {title} {Dynamics of anyon clusters in fractional quantum hall fluids},\ }\href {https://doi.org/10.1103/vgz6-z98r} {\bibfield  {journal} {\bibinfo  {journal} {Phys. Rev. B}\ }\textbf {\bibinfo {volume} {112}},\ \bibinfo {pages} {235112} (\bibinfo {year} {2025})},\ \bibinfo {note} {editors' Suggestion}\BibitemShut {NoStop}%
\bibitem [{\citenamefont {Gattu}\ and\ \citenamefont {Jain}(2025)}]{GattuJain2025MolecularAnyons}%
  \BibitemOpen
  \bibfield  {author} {\bibinfo {author} {\bibfnamefont {M.}~\bibnamefont {Gattu}}\ and\ \bibinfo {author} {\bibfnamefont {J.~K.}\ \bibnamefont {Jain}},\ }\bibfield  {title} {\bibinfo {title} {Molecular anyons in the fractional quantum hall effect},\ }\href {https://doi.org/10.1103/scl5-8pv6} {\bibfield  {journal} {\bibinfo  {journal} {Phys. Rev. Lett.}\ }\textbf {\bibinfo {volume} {135}},\ \bibinfo {pages} {236601} (\bibinfo {year} {2025})}\BibitemShut {NoStop}%
\bibitem [{\citenamefont {Li}\ \emph {et~al.}(2026)\citenamefont {Li}, \citenamefont {Nosov}, \citenamefont {Wang},\ and\ \citenamefont {Khalaf}}]{LiNosovWangKhalaf2026BoundStates}%
  \BibitemOpen
  \bibfield  {author} {\bibinfo {author} {\bibfnamefont {Q.}~\bibnamefont {Li}}, \bibinfo {author} {\bibfnamefont {P.~A.}\ \bibnamefont {Nosov}}, \bibinfo {author} {\bibfnamefont {T.}~\bibnamefont {Wang}},\ and\ \bibinfo {author} {\bibfnamefont {E.}~\bibnamefont {Khalaf}},\ }\href@noop {} {\bibinfo {title} {Bound states of anyons: a geometric quantization approach}} (\bibinfo {year} {2026}),\ \Eprint {https://arxiv.org/abs/2603.24701} {arXiv:2603.24701 [cond-mat.str-el]} \BibitemShut {NoStop}%
\bibitem [{\citenamefont {Lillieh{\"o}{\"o}k}\ \emph {et~al.}(1997)\citenamefont {Lillieh{\"o}{\"o}k}, \citenamefont {Lejnell}, \citenamefont {Karlhede},\ and\ \citenamefont {Sondhi}}]{LilliehookEtAl1997}%
  \BibitemOpen
  \bibfield  {author} {\bibinfo {author} {\bibfnamefont {D.}~\bibnamefont {Lillieh{\"o}{\"o}k}}, \bibinfo {author} {\bibfnamefont {K.}~\bibnamefont {Lejnell}}, \bibinfo {author} {\bibfnamefont {A.}~\bibnamefont {Karlhede}},\ and\ \bibinfo {author} {\bibfnamefont {S.~L.}\ \bibnamefont {Sondhi}},\ }\bibfield  {title} {\bibinfo {title} {Quantum hall skyrmions with higher topological charge},\ }\href {https://doi.org/10.1103/PhysRevB.56.6805} {\bibfield  {journal} {\bibinfo  {journal} {Phys. Rev. B}\ }\textbf {\bibinfo {volume} {56}},\ \bibinfo {pages} {6805} (\bibinfo {year} {1997})}\BibitemShut {NoStop}%
\end{thebibliography}%

\end{document}